\documentclass[showpacs,twocolumn,pra]{revtex4}

\usepackage{amsfonts}
\usepackage{amsmath}
\usepackage{amssymb}
\usepackage{graphicx}

\begin{document}

\title{Coherently enhanced Raman scattering in atomic vapor}
\author{Chun-Hua Yuan$^{1}$, L. Q. Chen$^{1}$, Jietai Jing$^{1}$, Z. Y. Ou$%
^{1,2}$, Weiping Zhang$^{1}$}
\affiliation{$^{1}$State Key Laboratory of Precision Spectroscopy, Department of Physics,
East China Normal University, Shanghai 200062, P. R. China\\
$^{2}$Department of Physics, Indiana University-Purdue University
Indianapolis, 402 North Blackford Street, Indianapolis, Indiana 46202, USA}
\date{\today }

\begin{abstract}
We present a scheme to obtain the coherently enhanced Raman scattering in
atomic vapor which is induced by a spin wave initially written by a weak
write laser. The enhancement of Raman scattering is dependent on the number
and the spatial distribution of the flipped atoms generated by the weak
write laser. Such an enhanced Raman scattering may have practical
applications in quantum information, nonlinear optics and laser spectroscopy
because of its simplicity.
\end{abstract}

\pacs{42.65.Dr, 42.50.Gy, 42.25.Bs}
\maketitle


\section{Introduction}

The Raman scattering technique \cite{Raman28} has enjoyed widespread
applications in molecular spectroscopy and light amplification. The
stimulated Raman scattering (SRS) process is well understood with a full
quantum theory \cite{Raymer81}. For Raman scattering in an atomic ensemble,
due to the limited number of atoms and limited interaction length, the
conversion efficiency of the pump field to Stokes laser is low and the SRS
regime is seldom reached. In order to obtain a high efficiency frequency
conversion, a seed Stokes field is usually injected into the atomic ensemble
and is used to generate the SRS in a short time.

\begin{figure}[tbp]
\centerline{\includegraphics[scale=0.55,angle=0]{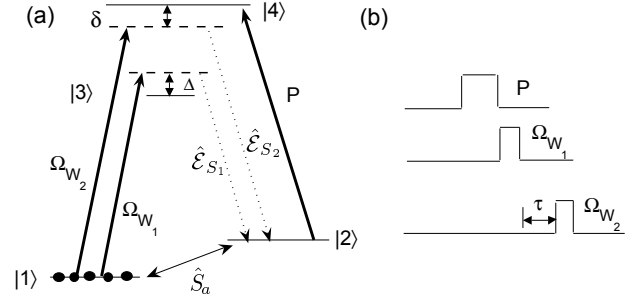}}
\caption{(a) The schematic diagram of four-level atoms. $P$: initial state
preparation laser; $\Omega_{W_1}$, $\Omega_{W_2}$: the Rabi frequencies of
write lasers; $\hat{\mathcal{E}}_{S_{1}}$, $\hat{\mathcal{E}}_{S_{2}}$: the
generated Raman Stokes signals; $\Delta$, $\protect\delta$: the detunings.
(b) The time sequence for different fields. $\protect\tau$: the delay time.}
\label{fig1}
\end{figure}

Another method to improve conversion efficiency is coherent anti-Stokes
Raman spectroscopy (CARS) \cite{Demtroder}, which utilizes the atomic
coherence built in stimulated Raman process to greatly enhance the
anti-Stokes component. This technique was developed into two regimes: one is
femtosecond adaptive spectroscopic techniques applied to CARS (FAST CARS),
which makes it easy to detect minute biological and chemical agents \cite%
{Scully02,Beadie,Sautenkov,Pestov}, another is enhanced four-wave mixing
(FWM) using electromagnetically induced transparency (EIT) \cite{Harris97}
for generation of non-classical correlations, non-classical photon-pairs
\cite%
{Lukin98,LV,Zibrov99,Wal,Kuzmich,Matsukevich104,Kolchin,Balic,Thompson,Du,Ooi,Yuan}
and single photons \cite{Chou,Polyakov} where the weak generated signals can
avoid resonant absorption due to EIT. These works can be understood as a
combined \textquotedblleft write-read\textquotedblright\ process. In the
write process, a spin wave (or an atomic coherence) is written in the atomic
ensemble. In the read process, the stored spin wave is retrieved by coherent
conversion from the atomic states into the anti-Stokes field. There are many
ways to prepare atomic coherence, such as EIT and stimulated Raman adiabatic
passage (STIRAP) \cite{Bergmann}. Jain \textit{et al.} \cite{Jain} achieved
a high-frequency conversion efficiency from 425 to 293 nm with the help of
an atomic coherence prepared via EIT on a Raman transition. Sautenkov
\textit{\ et al.} \cite{Sautenkov04} demonstrated the enhancement of
coherent anti-Stokes laser in Rb atomic vapor by a maximal atomic coherence
prepared by fractional STIRAP. Recently, our group observed an enhanced
Raman scattering (ERS) effect by the prebuilt atomic spin wave \cite{Chen09}%
. This effect can be understood as a \textquotedblleft
write-write\textquotedblright\ process shown in Fig. \ref{fig1}. In the
first step, the Stokes field $\hat{\mathcal{E}}_{S_{1}}$ is produced and a
spin wave is written into the atomic ensemble by the first write field $%
\Omega _{W_{1}}$, which is the same as the first step of the
\textquotedblleft write-read\textquotedblright\ process. In the second step
after some delay, differing from the CARS\ process, a second write field $%
\Omega _{W_{2}}$ is applied for another Raman scattering, thus another
Stokes field $\hat{\mathcal{E}}_{S_{2}}$ is generated.

In this article, we present a theoretical treatment for the observed ERS
effect. Here, we will show that the intensity of the second Stokes field $%
\hat{\mathcal{E}}_{S_{2}}$ can be enhanced compared to the case with no spin
wave prepared initially and depends on the intensity and the spatial
distribution of initially prepared atomic spin wave.

Our article is organized as follows. In Sec.~\ref{Raman}, we describe our
ERS based on the spin wave initially prepared by a weak write laser, and
give the intensities of the Stokes laser ($\hat{\mathcal{E}}_{S_{2}})$ in
the counterpropagation case and in the copropagation case. In Sec.~\ref%
{analysis}, we numerically calculate the intensities of the usual Raman
scattering (URS) and ERS, which shows that the intensities of ERS are larger
than that of URS and the intensity of counterpropagation case is larger that
of copropagation case. In Sec.~\ref{Discussion}, we discuss the ERS and URS.
Finally, we conclude with a summary of our results.

\section{Theoretical Model}

\label{Raman}

In order to explain why the Raman scattering effect shown in Fig.~1 can be
enhanced, we consider a three-level Raman system composed of states $%
|1\rangle $, $|2\rangle $, and $|4\rangle $, relating to the second write
field $W_{2}$. Due to the first write field $W_{1}$ is absent in the
four-level system, the process is the familiar URS. When two write fields $%
W_{1}$ and $W_{2}$ with respective Rabi frequencies $\Omega _{W_{1}}$ and $%
\Omega _{W_{2}}$ are driven on the atomic system according to the time
sequence shown in Fig.~\ref{fig1}(b), the intensity of Stokes field $\hat{%
\mathcal{E}}_{S_{2}}$ will be enhanced compared to the URS case due to the
coherence $\hat{\sigma}_{12}$ between states $|1\rangle $ and $|2\rangle $
is built by the first write field $W_{1}$ \cite{Jain,Oberst}. The
enhancement of the intensity of Stokes field $\hat{\mathcal{E}}_{S_{2}}$ is
also dependent on the spatial distribution of the flipped atoms generated by
the first write field $W_{1}$.

In the undepleted pump approximation, the interaction Hamiltonian in a
rotating wave frame is given by \cite{Raymer81,Andre}
\begin{eqnarray}
H &=&\frac{N}{V}\int d\mathbf{r}\{\delta \tilde{\sigma}_{44}(\mathbf{r}%
,t)-[\Omega _{W_{2}}e^{i\mathbf{k}_{W_{2}}\cdot \mathbf{r}}\tilde{\sigma}%
_{41}  \notag \\
&&+g_{2}\hat{\mathcal{E}}_{S_{2}}(\mathbf{r},t)e^{i\mathbf{k}_{S_{2}}\cdot
\mathbf{r}}\tilde{\sigma}_{42}+\text{H.c.}]\},
\end{eqnarray}%
where $\delta =(\omega _{4}-\omega _{1})-\omega _{W_{2}}$, $g_{2}$ is the
atom-field coupling constant, the Stokes electric field $\hat{E}_{S_{2}}(%
\mathbf{r},t)=\sqrt{\hbar \omega _{S_{2}}/2\varepsilon _{0}V}[\hat{\mathcal{E%
}}_{S_{2}}(\mathbf{r},t)e^{i(\mathbf{k}_{S_{2}}\cdot \mathbf{r}-\omega
_{S_{2}}t)}+$h.c.$]$, $N$ is the number of atoms, and $V$ is the
quantization volume. The collective atomic operators are $\tilde{\sigma}%
_{\mu \nu }(\mathbf{r},t)=1/N_{\mathbf{r}}\sum_{j=1}^{N_{\mathbf{r}}}\hat{%
\sigma}_{\mu \nu }^{j}(t)e^{-i\omega _{\mu \nu }t}$, where $\hat{\sigma}%
_{\mu \nu }^{j}(t)=|\mu \rangle _{jj}\langle \nu |$ is the transition
operator of the $j$th atom between states $|\mu \rangle $ and $|\nu \rangle $%
, and there is a small and macroscopic volume containing $N_{\mathbf{r}}$ $%
(N_{\mathbf{r}}\gg 1)$ atoms around position $\mathbf{r}$. The Stokes field
operator $\hat{\mathcal{E}}_{S_{2}}(\mathbf{r},t)$ obeys the wave equation
\begin{equation}
\ \left( \frac{\partial }{\partial t}+c\frac{\partial }{\partial z}-i\frac{c%
}{2k_{S_{2}}}\nabla _{\bot }^{2}\right) \hat{\mathcal{E}}_{S_{2}}(\mathbf{r}%
,t)=ig_{2}N\tilde{\sigma}_{24}e^{-i\mathbf{k}_{S_{2}}\cdot \mathbf{r}}.
\label{eq2}
\end{equation}

The equations of motion for the atomic operators in the Heisenberg picture
are
\begin{eqnarray}
\dot{\tilde{\sigma}}_{24} &=&-(\gamma _{2}+i\delta )\tilde{\sigma}%
_{24}+i\Omega _{W_{2}}e^{i\mathbf{k}_{W_{2}}\cdot \mathbf{r}}\tilde{\sigma}%
_{21}  \notag \\
&&+ig_{2}\hat{\mathcal{E}}_{S_{2}}(\tilde{\sigma}_{22}-\tilde{\sigma}%
_{44})e^{i\mathbf{k}_{S_{2}}\cdot \mathbf{r}}+\hat{F}_{24},  \label{eq3} \\
\ \dot{\tilde{\sigma}}_{21} &=&-\gamma _{s}\tilde{\sigma}_{21}+i\Omega
_{W_{2}}^{\ast }e^{-i\mathbf{k}_{W_{2}}\cdot \mathbf{r}}\tilde{\sigma}_{24}
\notag \\
&&-ig_{2}\hat{\mathcal{E}}_{S_{2}}e^{i\mathbf{k}_{S_{2}}\cdot \mathbf{r}}%
\tilde{\sigma}_{41}+\hat{F}_{21},  \label{eq4} \\
\ \ \dot{\tilde{\sigma}}_{41} &=&-(\gamma _{2}-i\delta )\tilde{\sigma}%
_{41}-i\Omega _{W_{2}}^{\ast }e^{-i\mathbf{k}_{W_{2}}\cdot \mathbf{r}}(%
\tilde{\sigma}_{11}-\tilde{\sigma}_{44})  \notag \\
&&-ig_{2}\hat{\mathcal{E}}_{S_{2}}^{\dagger }e^{-i\mathbf{k}_{S_{2}}\cdot
\mathbf{r}}\tilde{\sigma}_{21}+\hat{F}_{41},  \label{eq5}
\end{eqnarray}%
where $\gamma _{s}$ is the coherence ($\tilde{\sigma}_{12}$) decay rate and $%
\gamma _{2}$ is the decay rate of the excited state $|4\rangle $. $\hat{F}%
_{\mu \nu }$ are the Langevin noise operators for the atomic operator.

Due to large detuning $\delta \gg |\Omega _{W_{2}}|$, $\gamma _{2}$, we can
adiabatically eliminate the optical coherence $\tilde{\sigma}_{24}$ and $%
\tilde{\sigma}_{41}$. In the following, we consider the number of atoms
scattered to $|2\rangle $ at all times much smaller than $N$, and obtain
\begin{eqnarray}
\tilde{\sigma}_{24} &=&(1+i\frac{\gamma _{2}}{\delta })[\frac{\Omega _{W_{2}}%
}{\delta }e^{i\mathbf{k}_{W_{2}}\cdot \mathbf{r}}\tilde{\sigma}_{21}+\frac{%
g_{2}}{\delta }e^{i\mathbf{k}_{S_{2}}\cdot \mathbf{r}}\hat{\mathcal{E}}%
_{S_{2}}\tilde{\sigma}_{22}],~~  \label{eq6} \\
\tilde{\sigma}_{41} &=&\frac{\Omega _{W_{2}}^{\ast }}{\delta }e^{-i\mathbf{k}%
_{W_{2}}\cdot \mathbf{r}}\tilde{\sigma}_{11}+i\frac{\hat{F}_{41}}{\delta }.
\label{eq7}
\end{eqnarray}%
Using Eq.~(\ref{eq6}), we have the Stokes propagating equation
\begin{equation}
\left( \frac{\partial }{\partial t}+c\frac{\partial }{\partial z}-i\frac{c}{%
2k_{S_{2}}}\nabla _{\bot }^{2}\right) \hat{\mathcal{E}}_{S_{2}}(\mathbf{r}%
,t)=i\chi _{2}\hat{S}_{a_{2}}^{\dag },  \label{eq8}
\end{equation}%
where $\hat{S}_{a_{2}}^{\dag }=\sqrt{N}\tilde{\sigma}_{21}e^{i\Delta \mathbf{%
k}\cdot \mathbf{r}}$ is the creation operator of the atomic spin wave where
the wave vector $\Delta \mathbf{k}=\mathbf{k}_{W_{2}}-\mathbf{k}_{S_{2}}$,
and $\chi _{2}(\mathbf{r},t)=g_{2}\sqrt{N}\Omega _{W_{2}}(\mathbf{r}%
,t)/\delta $ is the coupling rate between the collective spin excitation $%
\hat{S}_{a_{2}}$ and the Stokes field $\hat{\mathcal{E}}_{S_{2}}$. Using
Eqs.~(\ref{eq6}) and (\ref{eq7}) we have
\begin{eqnarray}
\dot{\hat{{S}}}_{a_{2}}^{\dag } &=&-(\gamma _{S_{2}}-i\delta _{L_{2}})\hat{S}%
_{a_{2}}^{\dag }-i\chi _{2}^{\ast }\hat{\mathcal{E}}_{S_{2}}(\tilde{\sigma}%
_{11}-\tilde{\sigma}_{22})+\hat{F}_{S_{2}}^{\dagger },~~~  \label{eq9} \\
\mathcal{\dot{W}}(t) &\simeq &-\gamma _{2}^{\prime }(\tilde{\sigma}_{11}-%
\tilde{\sigma}_{22})-\gamma _{2}^{\prime },  \label{eq10}
\end{eqnarray}%
where $\mathcal{W}(t)=\tilde{\sigma}_{11}-\tilde{\sigma}_{22}$, $\gamma
_{S_{2}}=\gamma _{s}+\gamma _{2}^{\prime }$, $\gamma _{2}^{\prime }=\gamma
_{2}|\Omega _{W_{2}}|^{2}/\delta ^{2}$, and $\delta _{L_{2}}=|\Omega
_{W_{2}}|^{2}/\delta $ is the AC Stark shift. $\hat{F}_{S_{2}}^{\dagger }$
is the Langevin noise operator and $\langle \hat{F}_{S_{2}}(\mathbf{r},t)%
\hat{F}_{S_{2}}^{\dagger }(\mathbf{r}^{\prime },t^{\prime })\rangle =2\gamma
_{S_{2}}V\delta (\mathbf{r}-\mathbf{r}^{\prime })\delta (t-t^{\prime })$.
The population difference $\mathcal{W}(t)$\ is slowly varying in a time of
the order of an optical pulse due to large detuning $\delta \gg |\Omega
_{W_{2}}|$.

From Eq.~(\ref{eq10}), the atomic population difference is $\mathcal{W}(t)=%
\mathcal{W}(0)\exp (-\gamma _{2}^{\prime }t)+\exp (-\gamma _{2}^{\prime
}t)-1 $, where $\mathcal{W}(0)=\tilde{\sigma}_{11}(0)-\tilde{\sigma}_{22}(0)$%
. Therefore, $\tilde{\sigma}_{11}-\tilde{\sigma}_{22}$\ in Eq. (\ref{eq9})
can be replaced by $\mathcal{W}(t)$ which describes the small changes in
population difference. The equations of motion ($\ref{eq3}$)-($\ref{eq5}$)
are the same as that of URS case except the initial conditions are
different. In the case of URS \cite{Raymer81,Andre}, the atomic medium is
initially optically pumped to the ground state $|1\rangle $, then the
parameter $\mathcal{W}(0)=1$ and the other atomic operators are zero, but in
the case of ERS, the initial state is a superposition state of states $%
|1\rangle $ and $|2\rangle $, so$\ \langle \tilde{\sigma}_{11}\rangle \neq 0$%
, $\langle \tilde{\sigma}_{22}\rangle \neq 0$, and $\langle \tilde{\sigma}%
_{21}\rangle \neq 0$.

Assume the write field $\Omega _{W_{2}}$ corresponding to a focused beam and
the Fresnel number $\mathfrak{F}=A/\lambda _{S_{2}}L$ ($A$ is the
cross-sectional area, and $L$ is the cell length) is of the order of unity,
then only a single transverse spatial mode contributes strongly to emission
along the direction of the write field $\Omega _{W_{2}}$. Therefore the
above model can be simplified as a one-dimensional model, the propagating
quantized Stokes field $\hat{\mathcal{E}}_{S_{2}}$ obeys the equation of
motion \cite{Raymer81,Andre}
\begin{eqnarray}
(\frac{\partial }{\partial t}+c\frac{\partial }{\partial z})\hat{\mathcal{E}}%
_{S_{2}}(z,t) =i\chi _{2}\hat{S}_{a_{2}}^{\dag }, \\
\partial _{t}\hat{S}_{a_{2}}^{\dag } =-\Gamma _{S_{2}}\hat{S}_{a_{2}}^{\dag
}-i\mathcal{W}(t)\chi _{2}(t)\hat{\mathcal{E}}_{S_{2}}+\hat{F}%
_{S_{2}}^{\dagger },
\end{eqnarray}%
where $\Gamma _{S_{2}}=\gamma _{S_{2}}-i\delta _{L_{2}}$. Using the moving
frame $t^{\prime }=t-z/c$, $z^{\prime }=z$, the solution is \cite%
{Raymer81,Andre}
\begin{widetext}
\begin{eqnarray}
\hat{S}_{a_{2}}^{\dag }(z^{\prime },t^{\prime }) &=&e^{-\Gamma
_{2}(t^{\prime })}\hat{S}_{a_{2}}^{\dag }(z^{\prime
},0)-i\int_{0}^{t^{\prime }}\mathcal{W}(t^{\prime \prime })\chi
_{2}(t^{\prime \prime })e^{-[\Gamma _{2}(t^{\prime })-\Gamma _{2}(t^{\prime
\prime })]}H_{2}(z^{\prime },0,t^{\prime },t^{\prime \prime })\hat{\mathcal{E%
}}_{S_{2}}(0,t^{\prime \prime })dt^{\prime \prime }  \notag \\
&&+e^{-\Gamma _{2}(t^{\prime })}\int_{0}^{z^{\prime }}G_{S_{2}}(z^{\prime
},z^{\prime \prime },t^{\prime },0)\hat{S}_{a_{2}}^{\dagger }(z^{\prime
\prime },0)dz^{\prime \prime }+\int_{0}^{t^{\prime }}e^{-[\Gamma
_{2}(t^{\prime })-\Gamma _{2}(t^{\prime \prime })]}\hat{F}_{S_{2}}^{\dagger
}(z^{\prime },t^{\prime \prime })dt^{\prime \prime }  \notag \\
&&+\int_{0}^{t^{\prime }}e^{-[\Gamma _{2}(t^{\prime })-\Gamma _{2}(t^{\prime
\prime })]}\int_{0}^{z^{\prime }}G_{S_{2}}(z^{\prime },z^{\prime \prime
},t^{\prime },t^{\prime \prime })\hat{F}_{S_{2}}^{\dagger }(z^{\prime \prime
},t^{\prime \prime })dz^{\prime \prime }dt^{\prime \prime },
\label{spinwave} \\
\hat{\mathcal{E}}_{S_{2}}(z^{\prime },t^{\prime }) &=&\hat{\mathcal{E}}%
_{S_{2}}(0,t^{\prime })+\frac{\chi _{2}(t^{\prime })}{c}\int_{0}^{t^{\prime
}}\mathcal{W}(t^{\prime \prime })\chi _{2}(t^{\prime \prime })e^{-[\Gamma
_{2}(t^{\prime })-\Gamma _{2}(t^{\prime \prime })]}\hat{\mathcal{E}}%
_{S_{2}}(0,t^{\prime \prime })G_{e_{2}}(z^{\prime },0,t^{\prime },t^{\prime
\prime })dt^{\prime \prime }  \notag \\
&&+\frac{i\chi _{2}(t^{\prime })}{c}e^{-\Gamma _{2}(t^{\prime
})}\int_{0}^{z^{\prime }}H_{2}(z^{\prime },z^{\prime \prime },t^{\prime },0)%
\hat{S}_{a_{2}}^{\dagger }(z^{\prime \prime },0)dz^{\prime \prime }  \notag
\\
&&+\frac{i\chi _{2}(t^{\prime })}{c}\int_{0}^{t^{\prime }}e^{-[\Gamma
_{2}(t^{\prime })-\Gamma _{2}(t^{\prime \prime })]}\int_{0}^{z^{\prime
}}H_{2}(z^{\prime },z^{\prime \prime },t^{\prime },t^{\prime \prime })\hat{F}%
_{S_{2}}^{\dagger }(z^{\prime \prime },t^{\prime \prime })dz^{\prime \prime
}dt^{\prime \prime },  \label{Stokes}
\end{eqnarray}%
\end{widetext}where
\begin{eqnarray}
&&H_{2}(z^{\prime },z^{\prime \prime },t^{\prime },t^{\prime \prime
})=I_{0}(2\sqrt{[p_{2}(t^{\prime })-p_{2}(t^{\prime \prime })]\frac{%
z^{\prime }-z^{\prime \prime }}{c}}),  \notag \\
&&G_{e_{2}}(z^{\prime },z^{\prime \prime },t^{\prime },t^{\prime \prime })=%
\frac{c(z^{\prime }-z^{\prime \prime })}{p_{2}(t^{\prime })-p_{2}(t^{\prime
\prime })}G_{S_{2}}(z^{\prime },z^{\prime \prime },t^{\prime },t^{\prime
\prime }),  \notag \\
&&G_{S_{2}}(z^{\prime },z^{\prime \prime },t^{\prime },t^{\prime \prime })=%
\sqrt{\frac{p_{2}(t^{\prime })-p_{2}(t^{\prime \prime })}{c(z^{\prime
}-z^{\prime \prime })}}  \notag \\
&&~~~~~~~~~~\times I_{1}(2\sqrt{[p_{2}(t^{\prime })-p_{2}(t^{\prime \prime
})]\frac{z^{\prime }-z^{\prime \prime }}{c}}).
\end{eqnarray}%
Here, $\Gamma _{2}(t^{\prime })=\int_{0}^{t^{\prime }}\Gamma
_{S_{2}}(t^{\prime \prime })dt^{\prime \prime }$ and $p_{2}(t^{\prime
})=\int_{0}^{t^{\prime }}\mathcal{W}(t^{\prime \prime })\chi _{2}(t^{\prime
\prime })^{2}dt^{\prime \prime }$, and $I_{n}(x)$\ is the modified Bessel
function of the first kind of order $n$. This solution is also for
pencil-shaped atomic ensemble for URS case when $\mathcal{W}(0)=1$ ~\cite%
{Raymer81,Andre}.

In order to explain the ERS, we compare the intensity of the ERS with that
of the URS. Using Eq.~(\ref{Stokes}), the intensity at the end of the atomic
cell is given by%
\begin{widetext}
\begin{eqnarray}
I_{S_{2}}(t^{\prime }) &=&\frac{\hbar \omega _{S_{2}}c}{L}\langle \hat{%
\mathcal{E}}_{S_{2}}^{\dag }(L,t^{\prime })\hat{\mathcal{E}}%
_{S_{2}}(L,t^{\prime })\rangle   \notag \\
&=&\frac{\hbar \omega _{S_{2}}\chi _{2}^{2}(t^{\prime })}{cL}e^{-2\mathrm{Re}%
[\Gamma _{2}(t^{\prime })]}\int_{0}^{L}H_{2}(L,z^{\prime },t^{\prime
},0)\int_{0}^{L}H_{2}(L,z^{\prime \prime },t^{\prime },0)\langle \hat{S}%
_{a_{2}}(z^{\prime },0)\hat{S}_{a_{2}}^{\dag }(z^{\prime \prime },0)\rangle
dz^{\prime }dz^{\prime \prime }  \notag \\
&+&\frac{2\hbar \omega _{S_{2}}\chi _{2}^{2}(t^{\prime })L}{c}\gamma
_{S_{2}}\int_{0}^{t^{\prime }}e^{-2\mathrm{Re}[\Gamma _{2}(t^{\prime
})-\Gamma _{2}(t^{\prime \prime })]}[I_{0}^{2}(2\sqrt{\frac{[p_{2}(t^{\prime
})-p_{2}(t^{\prime \prime })]L}{c}})-I_{1}^{2}(2\sqrt{\frac{[p_{2}(t^{\prime
})-p_{2}(t^{\prime \prime })]L}{c}})]dt^{\prime \prime },~~
\label{Intensity1}
\end{eqnarray}
\end{widetext}where the first and second terms are from the third and fourth
terms in Eq.~(\ref{Stokes}). According to the commutation relation of spin
wave, the term $\langle \hat{S}_{a_{2}}(z^{\prime \prime },0)\hat{S}%
_{a_{2}}^{\dag }(z^{\prime },0)\rangle $ in Eq.~(\ref{Intensity1}) is
\begin{eqnarray}
\langle \hat{S}_{a_{2}}(z^{\prime },0)\hat{S}_{a_{2}}^{\dag }(z^{\prime
\prime },0)\rangle &=&L\delta (z^{\prime }-z^{\prime \prime })  \notag \\
&+&\langle \hat{S}_{a_{2}}^{\dag }(z^{\prime \prime },0)\hat{S}%
_{a_{2}}(z^{\prime },0)\rangle .  \label{commutation}
\end{eqnarray}%
For URS, $\langle \hat{S}_{a_{2}}^{\dag }(z^{\prime },0)\hat{S}%
_{a_{2}}(z^{\prime \prime },0)\rangle $ equals zero because no initial spin
wave is prepared, but $\langle \hat{S}_{a_{2}}^{\dag }(z^{\prime },0)\hat{S}%
_{a_{2}}(z^{\prime \prime },0)\rangle \neq 0$ for ERS. The term $\langle
\hat{S}_{a_{2}}^{\dag }(z^{\prime },0)\hat{S}_{a_{2}}(z^{\prime \prime
},0)\rangle $ due to the initially written spin wave, is the key for the
appearance of the ERS for the write field $\Omega _{W_{2}}$. For
convenience, we assume the write fields intensity being constant $[\Omega
_{W_{2}}(t^{\prime })=\Omega _{W_{2}}\theta (t^{\prime })]$, after being
switched on at $t^{\prime }=0$, so $\Gamma _{2}(t^{\prime })=\Gamma
_{S_{2}}t^{\prime }$, and $p_{2}(t^{\prime })=\eta (t^{\prime })\chi
_{2}^{2}t^{\prime }$ with $\eta (t^{\prime })=\mathcal{W}(0)(1-1/2\gamma
_{2}^{\prime }t^{\prime })-1/2\gamma _{2}^{\prime }t^{\prime }$.

In the case of URS (considering a Raman system composed of states $|1\rangle
$, $|2\rangle $, and $|4\rangle $), no initial Stokes field is externally
incident on the ensemble and no initial spin wave is written into the
ensemble, so the initial conditions for the Stokes field and the spin wave
are$\ $
\begin{equation}
\langle \hat{\mathcal{E}}_{S_{2}}^{\dag }(0,t^{\prime })\hat{\mathcal{E}}%
_{S_{2}}(0,t^{\prime })\rangle =0,\text{ \ }\langle \hat{S}_{a_{2}}^{\dag
}(z^{\prime },0)\hat{S}_{a_{2}}(z^{\prime },0)\rangle =0.  \label{initial1}
\end{equation}%
Then from Eqs.~(\ref{Intensity1})-(\ref{initial1}) the intensity of the
Stokes field in URS case is

\begin{eqnarray}
I_{S_{2}}(t^{\prime }) &=&\frac{\hbar \omega _{S_{2}}\chi _{2}^{2}L}{c}%
\{2\gamma _{S_{2}}\int_{0}^{t^{\prime }}dt^{\prime \prime }e^{-2\gamma
_{S_{2}}(t^{\prime }-t^{\prime \prime })}  \notag \\
&\times& \lbrack I_{0}(2\sqrt{Kp})^{2}-I_{1}(2\sqrt{Kp})^{2}]+e^{-2\gamma
_{S_{2}}t^{\prime }}  \notag \\
&\times& \lbrack I_{0}(2\sqrt{\frac{\eta (t^{\prime })\chi _{2}^{2}L}{c}%
t^{\prime }})^{2}-I_{1}(2\sqrt{\frac{\eta (t^{\prime })\chi _{2}^{2}L}{c}%
t^{\prime }})^{2}]\},  \notag \\
&&  \label{Intensity2}
\end{eqnarray}%
where $Kp=\frac{L}{c}\chi _{2}^{2}[\eta (t^{\prime })t^{\prime }-\eta
(t^{\prime \prime })t^{\prime \prime }]$, and $\mathcal{W}(0)=1$ in the case
of the URS.

In the case of ERS, like URS, no Stokes field is externally incident on the
ensemble, but the atomic ensemble contains a spin wave $\hat{S}%
_{a_{1}}^{\dag }$ written by the first write field $\Omega _{W_{1}}$, which
the form is $\hat{S}_{a_{1}}^{\dag }(z^{\prime },t_{1})=\sqrt{N}\tilde{\sigma%
}_{21}(z^{\prime },t_{1})e^{i\Delta kz}$, where $t_{1}$ is the duration of
the first write pulse \cite{annotate}. When the second write field $\Omega
_{W_{2}}$ is driven on the ensemble, the initial conditions are given by$\ $
\begin{equation}
\langle \hat{\mathcal{E}}_{S_{2}}^{\dag }(0,t^{\prime })\hat{\mathcal{E}}%
_{S_{2}}(0,t^{\prime })\rangle =0,\text{ }\langle \hat{S}_{a_{2}}^{\dag
}(z^{\prime },0)\hat{S}_{a_{2}}(z^{\prime },0)\rangle \neq 0.
\label{initial2}
\end{equation}%
Then from Eqs.~(\ref{Intensity1}), (\ref{commutation}) and (\ref{initial2}),
the intensity of the Stokes field in the ERS case is
\begin{eqnarray}
I_{S_{2}}^{\prime }(t^{\prime }) &=&I_{S_{\text{2-0}}}(t^{\prime })+I_{\text{%
add}}(t^{\prime }), \\
I_{\text{add}}(t^{\prime }) &=&\frac{\hbar \omega _{S_{2}}\chi _{2}^{2}}{cL}%
e^{-2\gamma _{S_{2}}t^{\prime }}\int_{0}^{L}H_{2}(L,z^{\prime },t^{\prime
},0)\int_{0}^{L}dz^{\prime }dz^{\prime \prime }  \notag \\
&\times &H_{2}(L,z^{\prime \prime },t^{\prime },0)\langle \hat{S}%
_{a_{2}}^{\dag }(z^{\prime },0)\hat{S}_{a_{2}}(z^{\prime \prime },0)\rangle .
\label{additional}
\end{eqnarray}%
where $I_{S_{\text{2-0}}}(t^{\prime })$ is the usual Raman intensity as Eq.~(%
\ref{Intensity2}) except here $\mathcal{W}(0)\neq 1$ in ERS case, and $I_{%
\text{add}}(t^{\prime })$ is the additional intensity generated by the
initially prepared spin wave.

\begin{figure}[tbp]
\centerline{\includegraphics[scale=0.55,angle=0]{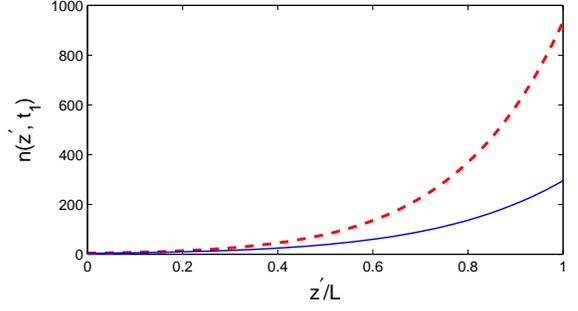}}
\caption{ (Color online) The number of flipped atoms versus the
dimensionless length $z^{\prime }/L$ for $\protect\zeta _{1}=6$ (solid line)
and $\protect\zeta _{1}=8$ (dashed line) with ${\protect\chi_{1}^{2}L}/c
\mathrm{Re}[\Gamma _{S_{1}}]=10$.}
\label{fig2}
\end{figure}

Next we investigate\ the additional intensity $I_{\text{add}}(t^{\prime })$.
We first analyze the term $\langle \hat{S}_{a_{2}}^{\dag }(z^{\prime },0)%
\hat{S}_{a_{2}}(z^{\prime \prime },0)\rangle $. We consider two different
propagation geometries: copropagating or counterpropagating of this two
write fields $\Omega _{W_{1}}$ and $\Omega _{W_{2}}$. For the copropagating
case, we have $\hat{S}_{a_{2}}(z^{\prime },0)=\hat{S}_{a_{1}}(z^{\prime
},t_{1})$, so
\begin{equation}
\langle \hat{S}_{a_{2}}^{\dag }(z^{\prime },0)\hat{S}_{a_{2}}(z^{\prime
\prime },0)\rangle =n(z^{\prime },t_{1})\delta (z^{\prime }-z^{\prime \prime
}),  \label{co}
\end{equation}%
and for two write fields counter-propagating case, we have $\hat{S}%
_{a_{2}}(z^{\prime },0)=\hat{S}_{a_{1}}(L-z^{\prime },t_{1})$, so
\begin{equation}
\langle \hat{S}_{a_{2}}^{\dag }(z^{\prime },0)\hat{S}_{a_{2}}(z^{\prime
\prime },0)\rangle =n(L-z^{\prime },t_{1})\delta (z^{\prime }-z^{\prime
\prime }),  \label{counter}
\end{equation}%
with
\begin{equation}
n(z^{\prime },t_{1})=\int_{0}^{\zeta _{1}}e^{-2c\zeta \mathrm{Re}[\Gamma
_{S_{1}}]/\chi _{1}^{2}L}I_{0}(2\sqrt{\zeta \frac{z^{\prime }}{L}}%
)^{2}d\zeta ,  \label{density}
\end{equation}%
where $\zeta _{1}=\chi _{1}^{2}Lt_{1}/c$, $\zeta $ ($\zeta =\chi
_{1}^{2}Lt/c $, $0\leq t\leq t_{1}$) are the dimensionless strengths \cite%
{annotate}, $\mathrm{Re}[Z]$ denotes the real part of a complex number $Z$,
and the coefficient ${\chi _{1}^{2}L}/c\mathrm{Re}[\Gamma _{S_{1}}]$ is the
order of optical depth. Figure~\ref{fig2} shows the spatial distribution of
the flipped-atom number prepared by the first write field $\Omega _{W_{1}}$
from Eq.~(\ref{density}), where the flipped density becomes larger toward
the end part of the atomic ensemble, and the flipped density increases with
increment of the dimensionless strength $\zeta _{1}$.

\begin{figure}[tbp]
\centerline{\includegraphics[scale=0.55,angle=0]{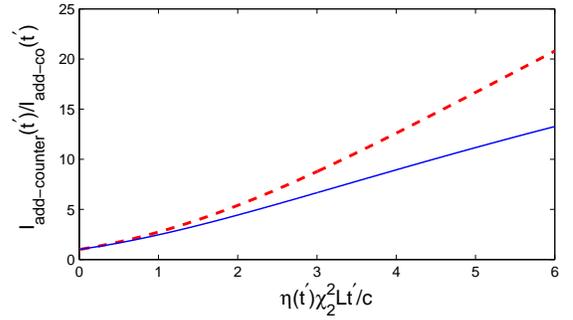}}
\caption{ (Color online) The ratio of the additional intensities $I_{\text{%
add-counter}}(t^{\prime })/I_{\text{add-co}}(t^{\prime })$ versus the
dimensionless strength $\protect\eta(t^{\prime })\protect\chi%
_2^{2}Lt^{\prime }/c $ for $\protect\zeta _{1}=6$ (solid line) and $\protect%
\zeta _{1}=8$ (dashed line) with ${\protect\chi_{1}^{2}L}/c\mathrm{Re}%
[\Gamma _{S_{1}}]=10$. }
\label{fig3}
\end{figure}

According to the co-propagating case and the counter-propagating case, the
additional intensities $I_{\text{add}}(t^{\prime })$ can be expressed as $I_{%
\text{add-co}}(t^{\prime })$ and $I_{\text{add-counter}}(t^{\prime })$. Thus
the intensities of Stokes field $\hat{\mathcal{E}}_{S_{2}}$ are written as
\begin{eqnarray}
I_{S_{\text{2-co}}}(t^{\prime }) &=&I_{S_{\text{2-0}}}(t^{\prime })+I_{\text{%
add-co}}(t^{\prime }), \\
I_{S_{\text{2-counter}}}(t^{\prime }) &=&I_{S_{\text{2-0}}}(t^{\prime })+I_{%
\text{add-counter}}(t^{\prime }).
\end{eqnarray}%
The spatial distribution will result in different ERS intensities $I_{S_{%
\text{2-co}}}(t^{\prime })$\ and $I_{S_{\text{2-counter}}}(t^{\prime })$ for
the copropagating case and the counterpropagating case, respectively. The
ratio $I_{\text{add-counter}}(t^{\prime })/I_{\text{add-co}}(t^{\prime })$
of the additional intensities $I_{\text{add-counter}}(t^{\prime })$ and $I_{%
\text{add-co}}(t^{\prime })$ is shown in Fig.~\ref{fig3}, from which we know
that the additional intensity of the counter-propagating case is much larger
that of the co-propagating case. The ratio increases with increment of the
dimensionless strength $\eta (t^{\prime })\chi _{2}^{2}Lt^{\prime }/c$, and
also increases with increment of the dimensionless strength $\zeta _{1}$.

\section{Numerical analysis}

\label{analysis} In this section, we will numerically calculate the
intensities of URS $I_{S_{2}}(t^{\prime })$, of ERS in the co-propagating
case$\ I_{\text{S}_{\text{2-co}}}(t^{\prime })$, and of ERS in the
counter-propagating case $I_{\text{S}_{\text{2-counter}}}(t^{\prime })$
using Gaussian-shaped write fields. Assume the write fields $\Omega _{W_{1}}$
and $\Omega _{W_{2}}$ have the following Gaussian shapes,
\begin{eqnarray}
\Omega _{W_{1}}(t^{\prime }) &=&\Omega _{W_{10}}[e^{-30(t^{\prime
}/T_{1}-0.5)^{2}}-e^{-7.5}], \\
\Omega _{W_{2}}(t^{\prime }) &=&\Omega _{W_{20}}[e^{-30(t^{\prime
}/T_{2}-0.5)^{2}}-e^{-7.5}],
\end{eqnarray}%
where $T_{1}$ and $T_{2}$\ are the pulse durations of write fields $W_{1}$
and $W_{2}$, respectively, and $\Omega _{W_{10}}$ and $\Omega _{W_{20}}$ are
the coefficients. From Eq.~(\ref{Intensity1}), we can obtain the
corresponding intensities for differently initial conditions. The parameters
$\Gamma _{2}(t^{\prime })$, $p_{2}(t^{\prime })$ in Eq.~(\ref{Intensity1})
are written as
\begin{eqnarray}
\Gamma _{2}(t^{\prime }) &=&\gamma _{s}t^{\prime }+\bar{\Gamma}%
_{2}\int_{0}^{t^{\prime }}[e^{-30(\frac{t^{\prime \prime }}{T_{2}}%
-0.5)^{2}}-e^{-7.5}]^{2}dt^{\prime \prime }, \\
p_{2}(t^{\prime }) &=&\bar{p}_{2}\int_{0}^{t^{\prime }}\mathcal{W}(t^{\prime
\prime })[e^{-30(\frac{t^{\prime \prime }}{T_{2}}%
-0.5)^{2}}-e^{-7.5}]^{2}dt^{\prime \prime },~~~~
\end{eqnarray}%
where $\bar{\Gamma}_{2}=\gamma _{2}|\Omega _{W_{20}}|^{2}/\delta
^{2}-i|\Omega _{W_{20}}|^{2}/\delta $, and$\ \bar{p}_{2}=(g_{2}\sqrt{N}%
\Omega _{W_{20}}/\delta )^{2}$. The term $\langle \hat{S}_{a_{2}}(z^{\prime
},0)\hat{S}_{a_{2}}^{\dagger }(z^{\prime \prime },0)\rangle $ in Eq.~(\ref%
{Intensity1}) for the two-write-fields copropagating case is calculated
according to
\begin{equation}
\langle \hat{S}_{a_{2}}(z^{\prime },0)\hat{S}_{a_{2}}^{\dagger }(z^{\prime
\prime },0)\rangle =\langle \hat{S}_{a_{1}}(z^{\prime \prime },T_{1})\hat{S}%
_{a_{1}}^{\dag }(z^{\prime \prime },T_{1})\rangle \delta (z^{\prime
}-z^{\prime \prime }),  \notag
\end{equation}%
where
\begin{widetext}
\begin{eqnarray}
&&\langle \hat{S}_{a_{1}}(z^{\prime \prime },T_{1})\hat{S}_{a_{1}}^{\dag
}(z^{\prime \prime },T_{1})\rangle =Le^{-2\mathrm{Re}[\Gamma
_{1}(T_{1})]}+2Le^{-2\mathrm{Re}[\Gamma _{1}(T_{1})]}\int_{0}^{z^{\prime
\prime }}G_{S_{1}}(z^{\prime \prime },z^{\prime \prime \prime
},T_{1},0)dz^{\prime \prime \prime }  \notag \\
&&+L^{2}e^{-2\mathrm{Re}[\Gamma _{1}(T_{1})]}\int_{0}^{z^{\prime \prime
}}G_{S_{1}}^{2}(z^{\prime \prime },z^{\prime \prime \prime
},T_{1},0)dz^{\prime \prime \prime }+4\gamma _{S_{1}}L\int_{0}^{T_{1}}e^{-2%
\mathrm{Re}[\Gamma _{1}(T_{1})-\Gamma _{1}(t^{\prime \prime
})]}\int_{0}^{z^{\prime \prime }}G_{S_{1}}(z^{\prime \prime },z^{\prime
\prime \prime },T_{1},0)dz^{\prime \prime \prime }dt^{\prime \prime }  \notag
\\
&&+2\gamma _{S_{1}}L\int_{0}^{T_{1}}e^{-2\mathrm{Re}[\Gamma
_{1}(T_{1})-\Gamma _{1}(t^{\prime \prime })]}dt^{\prime \prime }+2\gamma
_{S_{1}}L^{2}\int_{0}^{T_{1}}e^{-2\mathrm{Re}[\Gamma _{1}(T_{1})-\Gamma
_{1}(t^{\prime \prime })]}\int_{0}^{z^{\prime \prime
}}G_{S_{1}}^{2}(z^{\prime \prime },z^{\prime \prime \prime },T_{1},t^{\prime
\prime })dz^{\prime \prime \prime }dt^{\prime \prime },\label{eq32}
\end{eqnarray}%
\end{widetext}and%
\begin{eqnarray}
&&G_{S_{1}}(z^{\prime },z^{\prime \prime },t^{\prime },t^{\prime \prime })=%
\sqrt{\frac{p_{1}(t^{\prime })-p_{1}(t^{\prime \prime })}{c(z^{\prime
}-z^{\prime \prime })}}  \notag \\
&&~~~~~~~~~~~~~~~~~~~~~~~\times I_{1}(2\sqrt{[p_{1}(t^{\prime
})-p_{1}(t^{\prime \prime })]\frac{z^{\prime }-z^{\prime \prime }}{c}}),
\notag \\
&&\Gamma _{1}(t^{\prime })=\gamma _{s}t^{\prime }+\bar{\Gamma}%
_{1}\int_{0}^{t^{\prime }}[e^{-30(t^{\prime \prime
}/T_{1}-0.5)^{2}}-e^{-7.5}]^{2}dt^{\prime \prime },  \notag \\
&&p_{1}(t^{\prime })=\bar{p}_{1}\int_{0}^{t^{\prime }}[e^{-30(t^{\prime
\prime }/T_{1}-0.5)^{2}}-e^{-7.5}]^{2}dt^{\prime \prime }.
\end{eqnarray}%
Where $\bar{\Gamma}_{1}=\gamma _{1}|\Omega _{W_{10}}|^{2}/\Delta
^{2}-i|\Omega _{W_{10}}|^{2}/\Delta $, and $\bar{p}_{1}=(g_{1}\sqrt{N}\Omega
_{W_{10}}/\Delta )^{2}$. For two write fields counter-propagating case, $%
\langle \hat{S}_{a_{2}}(z^{\prime \prime },0)\hat{S}_{a_{2}}^{\dagger
}(z^{\prime \prime },0)\rangle =\langle \hat{S}_{a_{1}}(L-z^{\prime \prime
},T_{1})\hat{S}_{a_{1}}^{\dag }(L-z^{\prime \prime },T_{1})\rangle $.

\begin{figure}[tbp]
\centerline{\includegraphics[scale=0.6,angle=0]{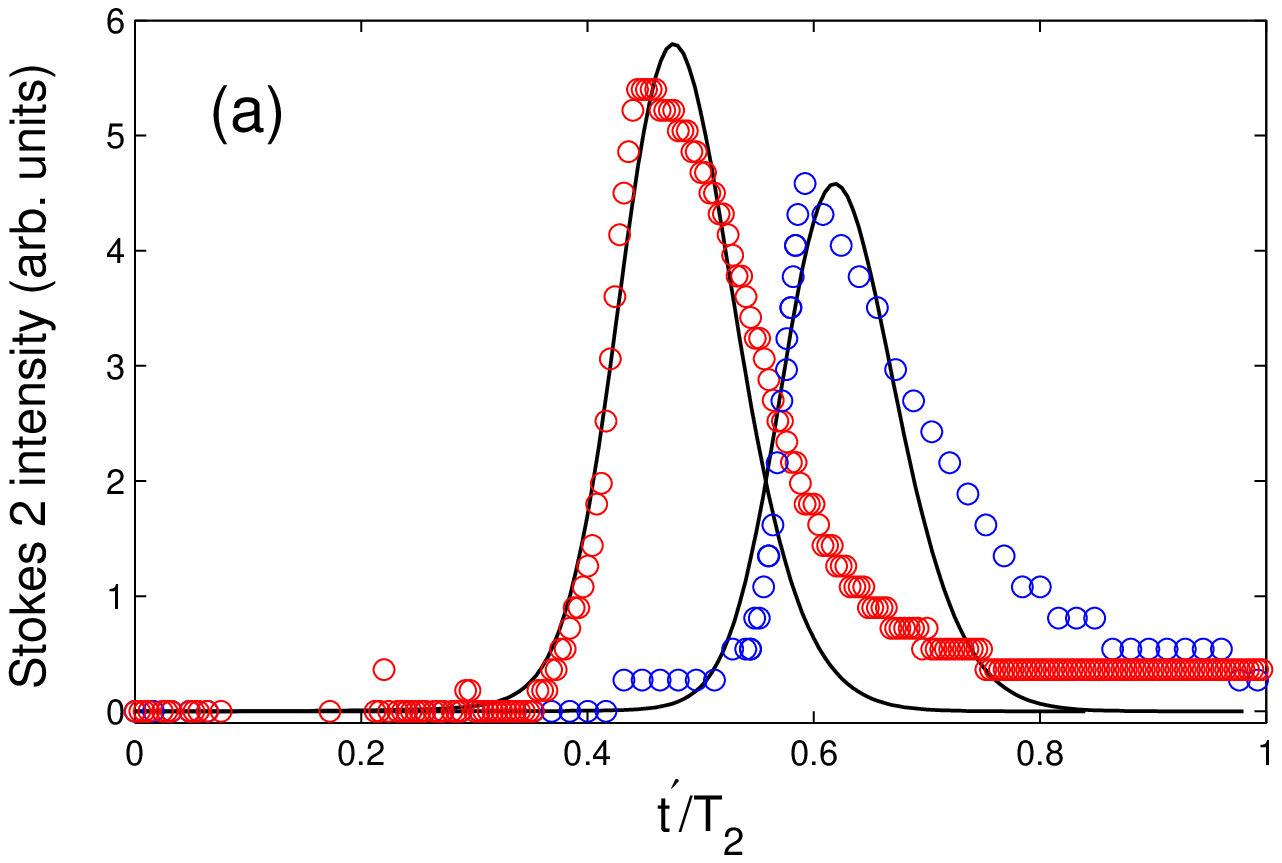}} %
\centerline{\includegraphics[scale=0.6,angle=0]{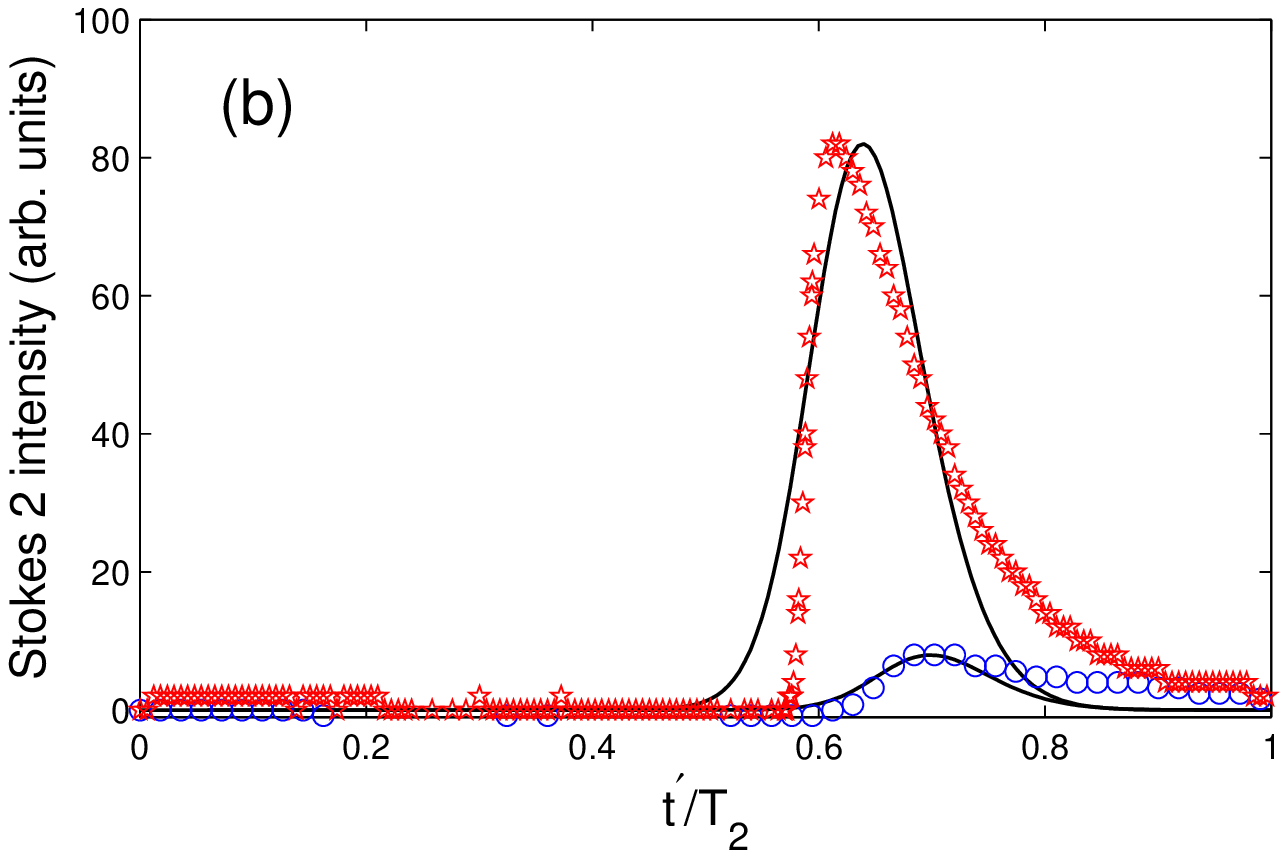}}
\caption{(Color online) The intensities of Stokes fields versus the
dimensionless time for (a) two write fields copropagating case and for (b)
two write fields counterpropagating case. The solid lines are from
theoretical results. The lines marked with ``o", ``square," and ``pentagram"
are from experimental data which describe the usual Raman ($\Omega
_{W_{1}}=0 $), the copropagation case of ERS, and the counterpropagation
case of ERS, respectively. The parameters are as follows: $\mathcal{W}%
(0)=0.99$, $Ng_1^2|\Omega_{10}|^{2}T_1L/(c\Delta^2)=8.5$, $%
|\Omega_2|/|\Omega_1|=1.56$, $\Delta=1.2$ GHz, $\protect\delta=1$ GHz, $%
\protect\gamma_s=10$ KHz, $\protect\gamma_1=2\protect\pi\times5.746$ MHz,
and $\protect\gamma_2=2\protect\pi\times6.605$ MHz. }
\label{fig4}
\end{figure}

Figure~\ref{fig4} numerically shows the intensities of three cases $%
I_{S_{2}}(t^{\prime })$, $I_{\text{S}_{\text{2-co}}}(t^{\prime })$, and $I_{%
\text{S}_{\text{2-counter}}}(t^{\prime })$ using Eqs.~(\ref{Intensity1}) and
(\ref{eq32}). In Fig.~\ref{fig4}, we choose the initial population
difference $\mathcal{W}(0)=0.99$, and the other experimental parameters are
the same as that were used in Ref.~\cite{Chen09}. When $Ng_{1}^{2}|\Omega
_{10}|^{2}T_{1}L/(c\Delta ^{2})=8.5$, the extent of enhancements agrees with
our recent experimental results \cite{Chen09}, where $Ng_{1}^{2}|\Omega
_{10}|^{2}T_{1}L/(c\Delta ^{2})$ is the order of the optical depth. When
only the write light $\Omega _{W_{2}}$ is turned on, a usual Raman
scattering occurs, as shown in the line marked with circles in Fig. \ref%
{fig4}. When the write fields $\Omega _{W_{1}}$ and $\Omega _{W_{2}}$ are
turned on according to the timing diagram in Fig.~\ref{fig1}(b) and are in
the copropagation configuration, an enhancement in $\hat{\mathcal{E}}%
_{S_{2}} $ occurs, where the experimental result corresponds to the line
marked with square and the theoretical numerical calculation corresponds to
the solid line in Fig.~\ref{fig4}(a). When the write fields $\Omega _{W_{1}}$
and $\Omega _{W_{2}}$ are in the counterpropagation configuration, a much
bigger enhancement effect in $\hat{\mathcal{E}}_{S_{2}}$ is shown in Fig.~%
\ref{fig4}(b), where the line marked with pentagram (the solid line) is from
the experimental (theoretical) data. From Fig.~\ref{fig4} it is easily seen
that the intensity of two write fields counterpropagating case is larger
than that of two write fields copropagating case. The reason why the
counter-propagating case has a larger enhancement effect than the
copropagating case is as follows. In the copropagating case, when the second
write field $\Omega _{W_{2}}$ enters the atomic medium, it first encounters
a very small number of the flipped atoms, which is equivalent to a small
seed for amplification. On the other hand, in the case of
counterpropagating, once the second write field $\Omega _{W_{2}}$ enters the
medium, it immediately encounters a maximum number of the flipped atoms and
starts to amplify it all the way through the atomic medium, which is
equivalent to a large seed for amplification.

\section{Discussion}

\label{Discussion}

In this section, let us compare the enhanced Raman process with the usual
Raman process. Assume the classical write field does not undergo depletion,
and then\ the usual Raman process is generated from spontaneous Raman
scattering at very small times to transient SRS at moderately small times,
and finally to steady-state SRS. That is to say, the spontaneous Stokes
scattering which is from the vacuum acts as the source term to generate SRS,
so the usual Raman process need a period of time to SRS and its phase is
random. But the enhanced Raman process is generated based on the prepared
spin wave and it directly and quickly to the SRS from the initiation. The
phase of the enhanced Stokes laser is from the spin wave and is nonrandom.
The atomic spin wave will act like an input seed to the Raman amplification
process in the same way as the input Stokes field would, so the intensity of
Stokes field $\hat{\mathcal{E}}_{S_{2}}$ is enhanced.

Our scheme can also be explained by the language of nonlinear optics. After
the first write laser is driven on the atomic ensemble, the atomic ensemble
is turned into a new medium within a certain coherence time, where the
nonlinear coefficient is larger compared to no initial coherence case. Then
when another write laser is driven on the atomic ensemble, an enhanced
Stokes field will occur.

\section{Conclusion}

In conclusion, we theoretically demonstrated an enhanced Raman effect, which
is another mechanism to realize SRS instantaneous and to increase the
conversion efficiency. The conversion efficiency of the second write laser
is high due to the initially prepared spin wave by the first write field,
and the enhancement of the Stokes field intensity in two-write-fields
counterpropagating case is much larger than that in the two-write-fields
copropagating case. The ERS is useful in quantum information, and in
nonlinear optics and to detect minute biological and chemical agents.



\begin{thebibliography}{99}
\bibitem{annotate} For an Raman system composed of states $|1\rangle $, $%
|2\rangle $, and $|3\rangle $, the Stokes field and spin wave are the same
as Eqs.~(\ref{spinwave}) and (\ref{Stokes}) except $\mathcal{W}=1$ and
subscripts $2\rightarrow 1$\ due to $\delta \rightarrow -\Delta $ and $%
\Omega _{W_{2}}\rightarrow \Omega _{W_{1}}$, so $\chi _{1}(\mathbf{r}%
,t)=g_{1}\sqrt{N}\Omega _{W_{1}}(\mathbf{r},t)/\Delta $ and $\Gamma
_{S_{1}}=\gamma _{S_{1}}-i\delta _{L_{1}}$ where $\delta _{L_{1}}=-|\Omega
_{W_{1}}|^{2}/\Delta $.
\end{thebibliography}
\end{document}